\documentclass[journal,comsoc]{IEEEtran}

\usepackage[T1]{fontenc}

\ifCLASSINFOpdf
\else
\fi

\usepackage{amsmath}
\usepackage{comment} %
\usepackage{cite}
\usepackage{graphicx}
\usepackage{epsfig,graphics,subfigure,psfrag,amsmath,amssymb}
\usepackage{booktabs}
\usepackage{amsfonts}
\usepackage{epstopdf}
\usepackage{amssymb}
\usepackage{array}
\usepackage{amsmath}
\usepackage{makecell}
\usepackage{multirow}
\usepackage[table,xcdraw]{xcolor}
\usepackage{colortbl}
\usepackage{tabularx}

\begin{document}
%
\title{Pinching-Antenna-Enabled ISAC: A Unified Architecture for Flexible Communication and Sensing}

\author{Yunshu Chen, Qing Xue, Chongjun Ouyang, Zhidu Li, Yi Wang, and Meng Hua

\thanks{Yunshu Chen, Qing Xue and Zhidu Li are with the School of Communications and Information Engineering, Chongqing University of Posts and Telecommunications, Chongqing 400065, China (e-mails: S240101060@stu.cqupt.edu.cn; xueq@cqupt.edu.cn; lizd@cqupt.edu.cn).}

\thanks{C. Ouyang is with the School of Electronic Engineering and Computer Science, Queen Mary University of London, London, E1 4NS, U.K. (e-mail: c.ouyang@qmul.ac.uk).}

\thanks{Yi Wang is with the School of Electronics and Information, Zhengzhou University of Aeronautics, Zhengzhou 450046, China (e-mail: yiwang@zua.edu.cn).}

\thanks{Meng Hua is with the Department of Electrical and Electronic Engineering, Imperial College London, SW7 2AZ London, U.K. (e-mail: m.hua@imperial.ac.uk).}

}

\maketitle

\begin{abstract}
Integrated sensing and communication (ISAC) is a cornerstone of sixth-generation (6G) networks, yet conventional fixed-antenna systems lack the spatial adaptability to cope with dynamic users and targets. The emerging pinching antenna (PA) offers a flexible, low-cost solution by dynamically reconfiguring radiation points along waveguides, introducing large-scale spatial degrees of freedom. This article develops a unified architectural perspective for PA-enabled ISAC. We first discuss the unique advantages of PAs over existing flexible solutions, and then propose a PA-enabled ISAC framework that accommodates both uplink and downlink communication while being compatible with passive and active sensing targets. Within this framework, we identify representative application scenarios, discuss major design challenges, and highlight critical enabling techniques. A numerical case study demonstrates how PA reconfigurability affects the communication-sensing rate trade-off. We also outline open issues to guide further PA-ISAC research for future 6G networks.
\end{abstract}

\IEEEpeerreviewmaketitle

\section{Introduction}

Emerging 6G use cases, such as intelligent transportation, immersive extended reality, and massive industrial Internet of Things (IoT), require wireless networks to support high-throughput data delivery and precise environmental sensing. Integrated sensing and communication (ISAC) has emerged as a key paradigm to meet this demand\cite{1}, by improving spectrum and hardware efficiency through unified signaling and processing. However, conventional ISAC systems mainly rely on fixed-antenna arrays, whose invariant positions limit their adaptability to dynamic and heterogeneous environments.

Recent advances in flexible antenna technologies, including reconfigurable intelligent surfaces (RIS), fluid antennas (FA), and movable antennas (MA), have introduced additional spatial degrees of freedom (DoFs) by reconfiguring wireless propagation environments or adjusting antenna positions\cite{3}. However, these approaches still face inherent limitations. For example, RIS suffer from the double-fading effect, while the movement ranges of FA and MA are typically restricted to the wavelength scale, which constrains their ability to effect large-scale coverage reshaping. Therefore, there is a pressing need for a flexible architecture that can overcome these issues while maintaining simplicity, scalability, and cost-effectiveness.

\begin{table*}[t]
\caption{Comparison of Key Features Among PA, RIS, and FA/MA Technologies.}
\label{tab_compare}
\setlength{\tabcolsep}{5mm}
\renewcommand{\arraystretch}{1.1}
\centering
\begin{tabular}{llll}
\hline
\textbf{Feature} & \textbf{PA} & \textbf{RIS} & \textbf{FA/MA} \\ \hline

\textbf{Deployment flexibility}
& Moderate
& High
& Low \\ \hline

\textbf{Control complexity}
& Moderate to low
& High
& Moderate \\ \hline

\textbf{Power consumption}
& Moderate
& Low
& Moderate \\ \hline

\textbf{Large-scale spatial relocation capability}
& Moderate
& Strong
& Weak \\ \hline

\textbf{RF-chain requirement}
& Moderate to low
& low
& Moderate \\ \hline

\end{tabular}
\end{table*}

Pinching antenna (PA) offers an attractive alternative. By placing controllable dielectric particles along a dielectric waveguide, PA systems can create reconfigurable radiation points that can establish strong line-of-sight (LoS) links at desired locations\cite{4,6}. Unlike conventional fixed arrays, PA systems enable larger-scale spatial reconfiguration along the waveguide with low hardware complexity, effectively decoupling antenna position from the physical aperture size. This inherent flexibility makes PA particularly attractive for ISAC scenarios that require adaptive coverage, directional target illumination, and reliable echo acquisition\cite{8}. As summarized in Table~I, among the three candidates, PA uniquely combines deployment flexibility, low control overhead, and radio frequency (RF)-chain economy, offering the most favorable trade-off for ISAC and thus emerging as a promising technological enabler.
While \cite{2} discussed related design methods and enabling techniques for PA-enabled ISAC, the associated network deployment architecture remains insufficiently investigated.

Considering the potential of PA in enhancing ISAC and the demand for flexible communication-sensing architectures in 6G, this article provides a unified architectural perspective on PA-enabled ISAC. First, we articulate the fundamental advantages of PA in ISAC, and propose a comprehensive PA-enabled ISAC framework supporting both uplink and downlink operation while being compatible with passive and active sensing targets, thereby demonstrating the architectural versatility of PA. Second, representative application scenarios, major design challenges, and key enabling techniques are analyzed for practical deployment. Third, a typical numerical case is presented to illustrate the impact of PA reconfigurability on the communication rate (CR)-sensing rate (SR) trade-off, revealing important insights into performance boundaries. Finally, we highlight several open issues to inspire future research on PA-empowered ISAC toward 6G networks.

\section{Fundamentals of PA-enabled ISAC Systems}

\subsection{Advantages of PA Employed in ISAC Systems}

Depending on the communication direction, ISAC transmission can be broadly categorized into downlink and uplink modes. In downlink, the base station (BS) actively transmits ISAC signals to simultaneously serve communication users and illuminate sensing targets. In uplink, user-transmitted signals can be opportunistically reused for sensing, enabling target detection without dedicated sensing transmissions from the BS. Both modes require flexible control of signal radiation, reception, and echo acquisition, which is difficult to meet with a fixed-antenna architecture.

PA offers a compelling set of advantages in this context. By placing dielectric particles on the surface of a waveguide, electromagnetic signals propagating inside the waveguide leak into free space at the corresponding positions, effectively creating reconfigurable radiation points\cite{9}. By properly configuring the pinching-point positions, PA can establish stronger LoS links in weak-coverage or blockage-prone regions, thereby improving both communication coverage and target sensing capability. Unlike conventional fixed-antenna arrays that rely solely on beamforming weights, PA systems can physically adjust effective antenna positions along the waveguide, thus providing an additional spatial DoFs for high-rate communication and high-accuracy sensing.

Furthermore, PA can support scalable and cost-effective deployment. Dielectric waveguides can be readily installed along walls, roads, or other industrial infrastructures to realize low-power long-distance coverage with minimal hardware overhead. For ISAC systems, this deployment flexibility not only enhances user service quality, but also facilitates multi-angle target observation, making PA particularly well-suited for complex and dynamic scenarios where both communication and sensing requirements evolve over time.

\subsection{PA-Enabled ISAC Framework}

Integrating PA with ISAC systems provides new insights for the joint design of communication and sensing, while offering more possibilities for system architecture design than conventional fixed-array solutions. To this end, this article separately presents the PA-enabled ISAC frameworks for downlink and uplink, and further considers passive and active sensing targets, as shown in Fig.~1. This four-quadrant design space captures the full diversity of real-world ISAC deployments.

\textbf{Downlink PA-enabled ISAC operation:} As illustrated in Fig.~1(a), the BS transmits ISAC signals to communication users through transmit PAs. For passive targets, the transmit PAs radiate ISAC signals that simultaneously serve users and illuminate targets. The users decode the embedded communication information, while the echo signals reflected by the passive targets are collected by the receive PAs for subsequent sensing processing. For active targets, they can autonomously generate sensing-related information and independently transmit sensing signals to the receive PAs, without relying on the reflection of incident BS signals. By exploiting the position reconfigurability of PAs, this architecture can dynamically adapt coverage and illumination patterns to varying user and target distributions. However, since the transmit and receive PAs share the same waveguide resources, strong transmitted signals may leak directly into receive PAs through spatial coupling, causing self-interference that may overwhelm weak target echoes and degrade detection accuracy\cite{7}. The partitioning and spatial distribution of the transmit and receive PAs therefore play a critical role in overall system performance.

\textbf{Uplink PA-enabled ISAC operation:}
In Fig.~1(b), communication users actively transmit uplink signals to the receive PAs. For passive targets, the user-transmitted communication signals simultaneously illuminate the targets during propagation and reach the receive PAs after reflection. In this case, the BS receives both the direct uplink communication signals and target echoes, reusing communication signals for target detection without dedicated sensing transmissions. For active targets, both the targets and communication users transmit signals to the receive PAs, where the user signals carry communication data, while the active sensing signals convey sensing-related information. In the uplink mode, the BS mainly performs signal reception, and only receive PAs are typically deployed along the waveguide. Therefore, self-interference caused by transmit-to-receive leakage can be avoided, while both transmit power consumption and transceiver hardware complexity can also be reduced. However, the sensing performance in this mode largely depends on the transmit power of the users or active targets, and the target echoes may be difficult to reliably detect.

Overall, the flexible configuration of PA functions and positions enables the proposed framework to adapt transmission and reception strategies to  diverse communication and sensing requirements.

\begin{figure*}
	\centering
	\includegraphics[width=0.85\textwidth]{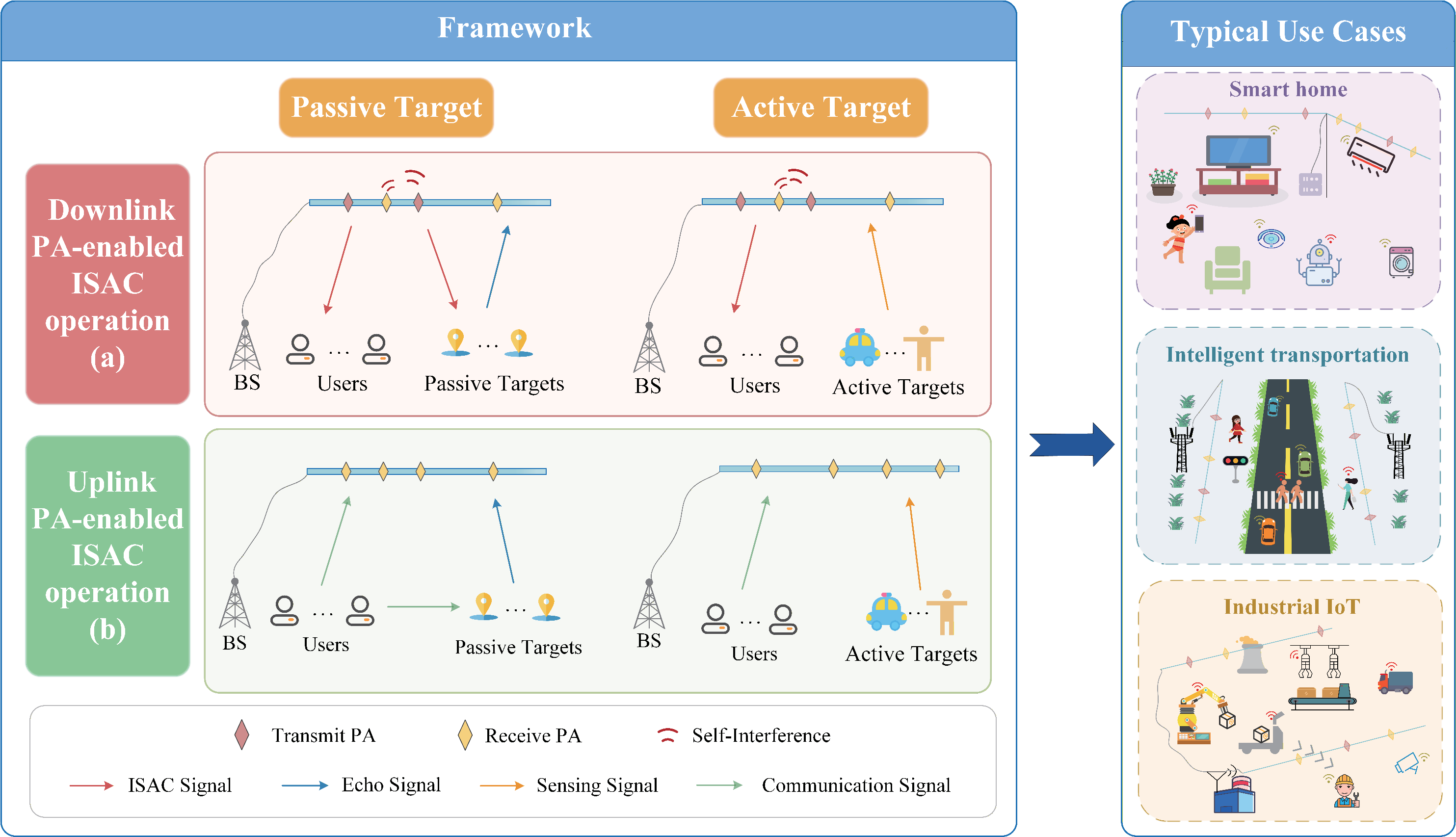}
	\caption{Illustration of the proposed PA-enabled ISAC framework and its typical application scenarios.}
	\label{fig:1}
\end{figure*}

Several typical application scenarios shown in Fig.~1 are discussed to further illustrate the potential value of PA-enabled ISAC.
In smart homes, PA-enabled ISAC can improve reliable connectivity and fine-grained sensing for intelligent terminals, wearable devices, and indoor targets by deploying dielectric waveguides along walls or ceilings, thereby enhancing downlink coverage, illuminating passive targets, and reusing uplink signals for activity sensing and status monitoring. In intelligent transportation, waveguides installed along roadsides, tunnels, or station infrastructures can provide continuous signal coverage for vehicles and pedestrians, while uplink signals from vehicles or active sensing devices can be collected for multi-angle observation and cooperative sensing across road segments. In industrial IoT, PA-enabled ISAC can use waveguides installed along production lines or warehouse facilities to concentrate downlink communication and sensing resources on specific working areas, while exploiting uplink signals from sensors or robots for equipment monitoring and production-line inspection.

\section{System Design for PA-Enabled ISAC}

Building on the proposed framework and its representative use cases, this section turns to the practical realization of PA-enabled ISAC systems. We first identify the fundamental design challenges that arise from the unique physical-layer characteristics of PA, and then discuss critical enabling techniques to address these challenges.

\subsection{Major Design Challenges}

Although the proposed PA-enabled ISAC framework provides flexible support for different communication directions and sensing target types, its practical implementation still faces several fundamental challenges.

\textit{Channel State Information (CSI) Acquisition Under PA Reconfiguration.} Reliable CSI is a prerequisite for accurately coordinating communication and sensing in ISAC systems. In PA-enabled ISAC, however, the channel response varies with the pinching-point positions and the propagation characteristics inside the waveguide. This means that when the PA configuration changes frequently, previously acquired CSI may no longer match the current PA layout, requiring repeated estimation of position-dependent channels and causing frequent CSI updates and large training overhead. In addition, communication users, passive targets, and active targets have different channel propagation structures. These factors make CSI highly position-dependent and difficult to characterize in a unified manner, which further affects beamforming and sensing parameter estimation.

\textit{Asymmetric PA Coverage for Communication and Sensing.} Communication users and sensing targets are generally located in different regions and have distinct service requirements, leading to different PA deployment requirements. For communication, PAs should be positioned to establish strong links with users and improve the coverage of weak-signal areas. For sensing, PA placement should facilitate target illumination and echo acquisition, and the favorable positions of transmit and receive PAs may be different. This asymmetry becomes more pronounced between the downlink and uplink modes. In the downlink mode, the BS can actively configure transmit PA positions to control user coverage and target illumination. In the uplink mode, however, sensing coverage largely depends on the locations and transmit powers of communication users or active targets, making target illumination and echo reception less controllable. Therefore, a PA configuration favorable for communication may not necessarily provide sufficient sensing performance, making it challenging to determine PA functions, positions, and beam patterns.

\textit{Self-Interference and Signal Leakage.} In the downlink mode, transmit and receive PAs simultaneously operate on the same waveguide. Strong transmitted signals can leak into the receive PAs through waveguide coupling or free-space propagation, resulting in severe self-interference. Since target echoes are usually much weaker than the transmitted signals, such leakage may overwhelm the desired echoes and degrade target detection and parameter estimation accuracy. Moreover, the interference level depends on the partitioning and spatial distribution of transmit and receive PAs. Although conventional transmit-to-receive self-interference can generally be avoided in the uplink mode, signals captured by one PA in the receive waveguide may be re-radiated by other PAs during propagation, further increasing the complexity of echo acquisition. Meanwhile, communication signals, passive-target echoes, and active-target sensing signals may still overlap at the receive PAs. Consequently, effective echo acquisition and interference suppression are essential.

\textit{Dynamic Reconfiguration and Practical Implementation.} The appropriate PA functions and positions vary with the propagation environment. When users or targets move, the system needs to rapidly update PA positions and signal processing strategies. However, the joint optimization of PA placement, beamforming, and resource allocation is typically high-dimensional and non-convex, making real-time optimization difficult. Practical PA systems are also affected by waveguide loss, positioning errors, hardware response latency, and imperfect calibration. Frequent PA adjustment may further introduce control overhead and reduce system stability. Deploying PA-enabled ISAC in dynamic environments requires agile and reliable implementation methods.

\subsection{Critical Techniques}

To address the above design challenges and fully exploit the advantages of PA-enabled ISAC systems in large-scale dynamic scenarios, several core enabling techniques must be developed, as discussed below.

\textit{Waveguide-Aware ISAC Channel and Echo Modeling.}
Accurate channel and echo models are fundamental to the design of PA-enabled ISAC systems. Unlike conventional fixed-array systems, the effective channel depends not only on free-space propagation, but also on waveguide propagation, pinching-point positions and the in-waveguide loss\cite{10}. For passive targets, the sensing channel further involves the cascaded propagation path from the signal source to the target and then to the receive PAs, whereas the sensing channel of active targets mainly depends on their direct links to the receive PAs. These heterogeneous propagation mechanisms complicate CSI acquisition, echo modeling, and sensing parameter extraction. For mobile or non-cooperative targets, sensing-assisted methods can infer their positions and motion states from echo signals and subsequently reconstruct or track the corresponding CSI\cite{11}. Therefore, efficient pilot design, position-aware channel estimation, and low-overhead CSI tracking are required. Model-based and data-driven approaches may also be combined to capture waveguide characteristics, target reflections, and hardware imperfections with manageable complexity.

\textit{Joint Pinching-Point Placement and ISAC Beam Synthesis.}
Flexible pinching-point placement provides PA-enabled ISAC systems with additional spatial control over signal radiation and reception, which is essential for addressing asymmetric communication and sensing coverage. Since pinching-point positions directly affect propagation phases, communication coverage, target illumination, and echo acquisition, they should be jointly designed with transmit/receive beamforming, PA array partitioning, and power allocation. Unlike conventional fixed-array beamforming, PA-enabled ISAC can synthesize beams through both signal weighting and physical adjustment of pinching-point positions, thereby realizing pinching beamforming\cite{12}. Beamforming enables the signals radiated or received by multiple PAs to be coherently combined, thereby directing energy toward communication users while maintaining sufficient sensing capability. Meanwhile, power allocation determines how the available transmit power is distributed among different waveguides, users, and sensing targets\cite{13}, which is important for improving weak communication links and satisfying sensing requirements. However, such joint design generally leads to a high-dimensional and non-convex problem, requiring low-complexity optimization or learning-driven methods for large-scale practical applications.

\textit{PA-Based Echo Acquisition and Tx/Rx Isolation for ISAC.}
The quality of echo acquisition directly affects the reliability of target detection, localization, and parameter estimation. In PA-enabled ISAC, the positions and spatial distribution of receive PAs determine the strength and diversity of the collected echoes, and should therefore be optimized according to the target region and sensing task to enhance weak-echo acquisition and mitigate the impact of signal leakage and self-interference. For passive targets, weak echoes need to be extracted from noise, clutter, communication signals, and strong transmit leakage. For active targets, sensing signals generated by the targets may overlap with uplink communication signals at the receive PAs. Accordingly, PA-based echo acquisition should be combined with receive beamforming, signal separation, and parameter estimation methods, such as delay, angle, Doppler, and position estimation.
In the downlink mode, the induced self-interference may be mitigated by attempting robust receiver designs, which suppress interference while preserving useful sensing echoes and communication signals. Meanwhile, segmented waveguide architectures isolate individual waveguide segments, providing a potential solution to the inter-antenna radiation caused by signal re-leaked in receive waveguides\cite{14}.

\textit{Fast Control, Calibration, and Robust ISAC Implementation.}
To support dynamic PA-enabled ISAC operation and implementation, practical systems require efficient control mechanisms to update PA positions with low latency. When the environment changes, hierarchical control and codebook-based configuration methods can be adopted to quickly select candidate PA layouts and beam patterns. By exploiting historical configurations and environmental features, warm-start optimization and learning-assisted decision methods can further reduce online computational overhead. In addition, calibration plays an important role in maintaining consistency between the designed and actual PA configurations. Factors such as fabrication errors and coupling effects may cause deviations in the expected radiation and reception patterns. Therefore, practical implementations should incorporate calibration procedures and robust design strategies that can cope with imperfect CSI and PA positioning errors. These techniques are essential for ensuring stable communication and sensing performance under dynamic and non-ideal deployment conditions.

\section{Numerical Case Study}

\begin{figure}
   \centering
   \includegraphics[width=0.95\columnwidth]{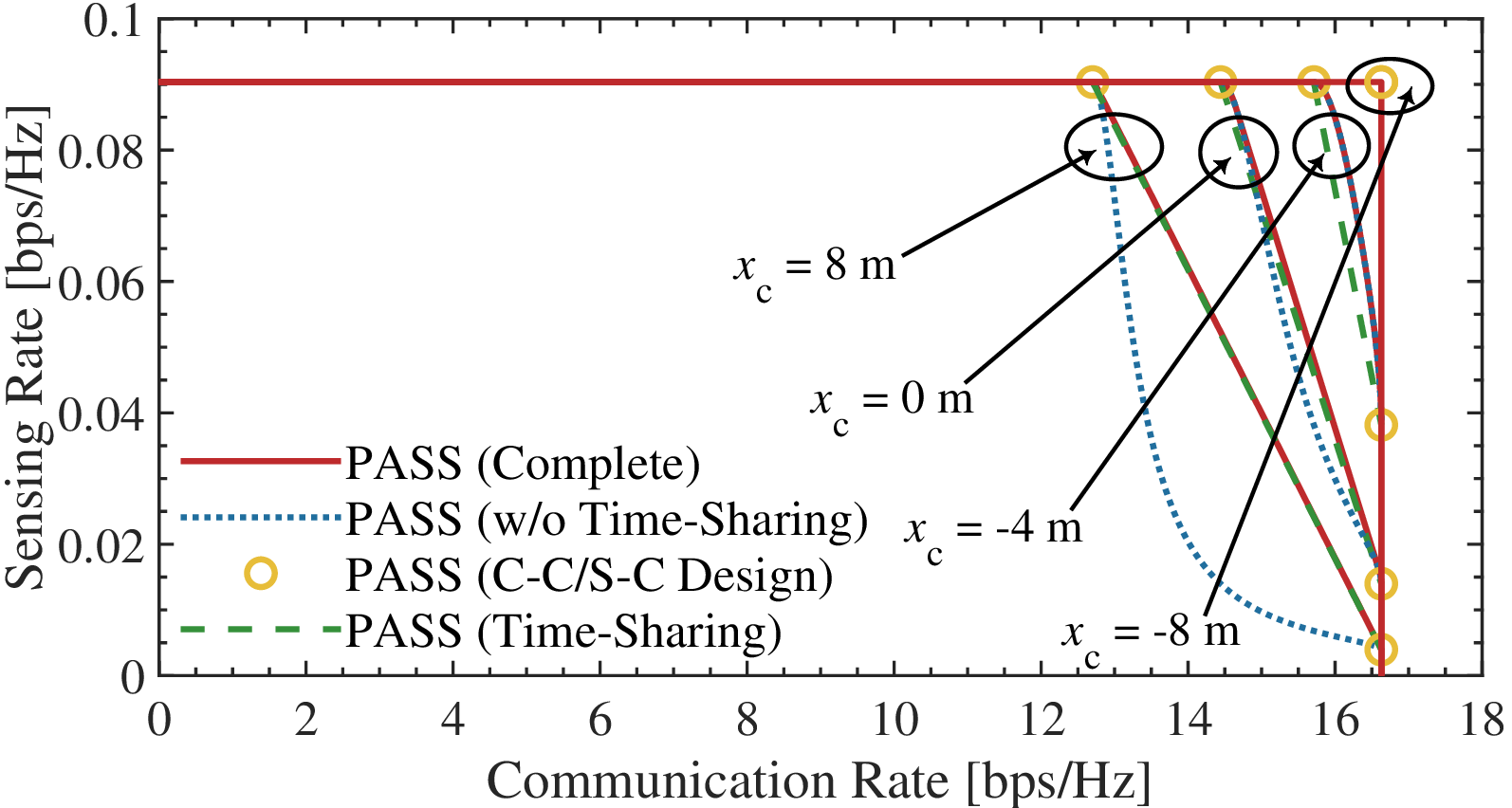}
   \caption{Instantaneous CR-SR rate region of the single-pinch PA-enabled ISAC system. The detailed simulation setup is given in \cite{5}.}
   \label{fig:2}
\end{figure}

\begin{figure}
   \centering
   \includegraphics[width=0.92\columnwidth]{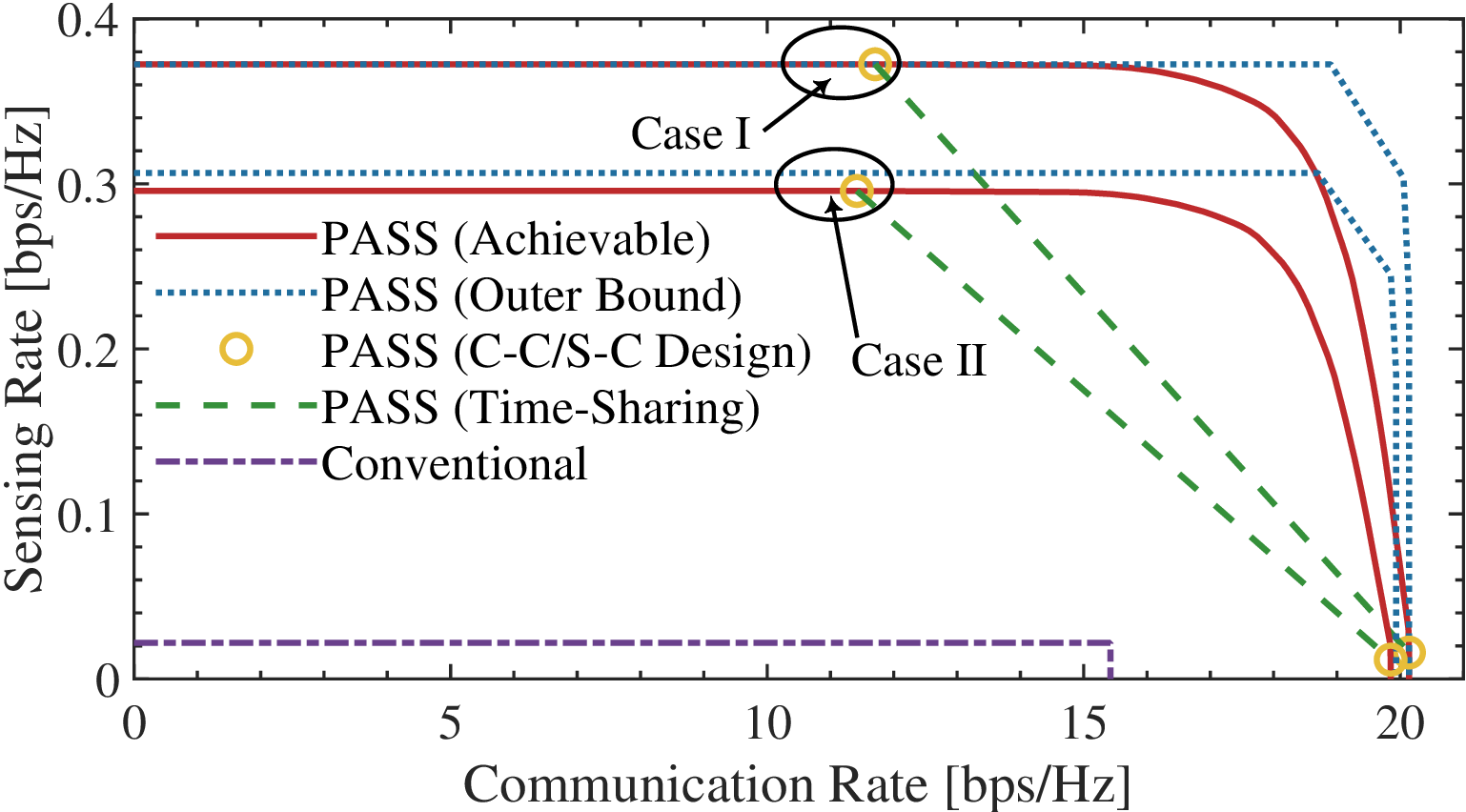}
   \caption{Average CR-SR rate region of the multiple-pinch PA-enabled ISAC system. The detailed simulation setup is given in \cite{5}.}
   \label{fig:3}
\end{figure}

To quantify the performance implications of PA reconfigurability in ISAC systems, this section presents a representative numerical case study. This case considers a downlink configuration with separate transmit and receive waveguides, where the system simultaneously serves a single user and senses a single target. By characterizing the achievable rate region between the CR and the SR, the impact of PA on the communication-sensing performance trade-off of ISAC systems is further evaluated.

\subsection{Single-Pinch Instantaneous Rate-Region}

We first examine a basic PA-enabled ISAC scenario with one transmit PA and one receive PA. In this setup, the in-waveguide loss is ignored, the target location is fixed, and the receive PA is aligned with the projection of the target along the waveguide direction. The user and the target are placed symmetrically with respect to the $x$-axis in the vertical coordinate, while only the user's horizontal coordinate varies. The numerical results in Fig.~2 illustrate how the relative geometry between the user and the target affects the CR-SR trade-off, as detailed in\cite[Fig.~5]{5}. To highlight the instantaneous rate regions achieved by PA under different user locations, the complete rate region, the achievable rate region, and the communications-centric (C-C)/sensing-centric (S-C) time-sharing region are compared.

\textit{Key observations.} When the user moves away from the target along the horizontal direction, the transmit pinching-point locations favorable for communication and sensing become different, resulting in an evident CR-SR trade-off. In this case, switching between the C-C and S-C designs can provide a reasonable approximation to the complete rate region. When the user is close to the target, a single pinching-point location can simultaneously satisfy both communication and sensing requirements. The achievable rate region thus becomes more rectangular, indicating that the conflict between CR and SR is weakened. These results show that, even in the single-pinch case, PA position optimization can significantly affect the achievable performance boundary of the system.

\subsection{Multiple-Pinch Average Rate-Region}

Fig.~3 further compares the average CR-SR rate region of the PA-enabled ISAC system, as illustrated in\cite[Fig.~9]{5}. In this multi-PA scenario, multiple PAs are deployed on the transmit waveguide, while a single receive PA remains aligned with the target. The user and target locations are assumed to follow uniform distributions. The figure compares the achievable rate region, the theoretical outer bound, the C-C/S-C time-sharing region, and the rate region of the conventional fixed-antenna system, under both the ideal waveguide case (Case I) and the in-waveguide loss case (Case II).

\textit{Key observations.} Under both waveguide conditions, the PA-enabled system achieves a rate region significantly larger than that of the conventional fixed-antenna system. This indicates that multiple PAs can provide more spatial DoFs, enabling more flexible radiation patterns and spatial energy distributions across varying user and target locations. Notably, time sharing only between the C-C and S-C designs is insufficient to approximate the complete rate region. This is because, in the multi-PA scenario, the coupling between pinching-point locations and beamforming becomes more complicated, and simple C-C/S-C switching cannot fully exploit the gain brought by PA reconfigurability. This demonstrates the necessity of joint optimization for balancing communication and sensing performance.

\textit{Summary.} The above numerical results reveal two key insights. First, PA position reconfigurability directly influences the achievable CR-SR frontier, with the impact being most pronounced when communication and sensing geometries diverge. Second, the benefits of PA scale with the number of pinching points, but fully realizing these gains requires sophisticated joint optimization that goes beyond simple mode switching. These findings validate the practical relevance of the PA-enabled ISAC framework and motivate the development of the enabling techniques discussed in Section III.

\section{Open Issues}

Despite the significant potential of PA-enabled ISAC systems for future large-scale 6G networks, their practical realization remains in its infancy, with many unresolved issues, bringing new opportunities and challenges for future exploration.

\emph{PA-Assisted Multi-Cell Synchronization and Distributed Sensing Fusion.} In large-scale PA-enabled ISAC deployments, multiple cells may need to cooperate to provide continuous communication coverage and wide-area sensing. However, distributed sensing relies on accurate time, frequency, and phase synchronization. Synchronization errors may lead to inconsistent delay, Doppler, and angle measurements, thereby degrading target localization accuracy.
Moreover, different cells may observe the same target from different angles and under different channel and interference conditions. The collected sensing information may have different resolutions, reliabilities, and update rates, making distributed sensing fusion difficult. Therefore, synchronization-aware sensing fusion methods, such as efficient cooperative tracking, distributed inference, and learning-assisted fusion mechanisms, are promising directions for improving sensing robustness in multi-cell PA-enabled ISAC networks.

\emph{PA Position Control Latency and Hardware Response.} Although fast control methods can reduce the computational latency of PA reconfiguration, the physical response of practical PA hardware may still limit real-time system operation. The mechanical switching of pinching elements, electromagnetic settling time, and control-circuit response delay may cause the optimized PA configuration to lag behind the actual channel and sensing environment, leading to mismatched beam patterns and degraded performance. Possible solutions include predictive PA control, hardware-software co-design, response-aware beamforming, and lightweight feedback mechanisms to compensate for configuration mismatches.

\emph{PA-Enabled Wireless Physical Neural Networks.} PA-enabled wireless physical neural networks represent a promising direction for integrating wireless propagation, sensing, and computation. Wireless physical neural networks aim to exploit the propagation environment itself as a physical computing medium.
For PA-enabled ISAC systems, such a mechanism is particularly attractive. By properly configuring pinching-point positions, the wireless channel can be shaped to enhance task-relevant features, such as target direction, motion patterns, or environmental changes, before digital processing. This may reduce the computational burden at the BS and enable low-latency sensing. However, PA-enabled wireless physical neural networks are still at an early stage. Issues such as establishing differentiable models linking PA configurations with neural computation and training physical parameters under hardware constraints remain key problems to be addressed.

\section{Conclusion}

In this article, we develop a unified architectural perspective for PA-enabled ISAC. We first introduced the advantages of PA in ISAC systems and proposed a PA-enabled ISAC framework supporting both uplink and downlink communication. Then, representative application scenarios of this framework were discussed, together with its major design challenges and key enabling techniques. Through numerical case studies, we further demonstrated the impact of PA on the CR-SR trade-off in ISAC systems. Finally, several open research directions were outlined, providing useful insights for the future development of PA-enabled ISAC toward 6G networks.

\ifCLASSOPTIONcaptionsoff
  \newpage
\fi

\bibliographystyle{IEEEtran}
\bibliography{reference}

@ARTICLE{1,
  author={Liu, Fan and Cui, Yuanhao and Masouros, Christos and Xu, Jie and Han, Tony Xiao and Eldar, Yonina C. and Buzzi, Stefano},
  journal={IEEE J. Sel. Areas Commun.},
  title={Integrated Sensing and Communications: Toward Dual-Functional Wireless Networks for {6G} and Beyond},
  year={2022},
  volume={40},
  number={6},
  month= jun,
  pages={1728-1767},
  doi={10.1109/JSAC.2022.3156632}}

@ARTICLE{2,
  author={Qin, Yunhui and Fu, Yaru and Liang, Yan and Cao, Difei and Zhang, Haijun},
  journal={IEEE Commun. Mag.},
  title={Pinching-Antenna Systems {(PASS)}: Paving the Way for {6G} {ISAC} Innovations},
  year={2026},
  volume={64},
  number={5},
  month= may,
  pages={212-218},
  doi={10.1109/MCOM.001.2500439}}

@ARTICLE{3,
  author={Yu, Xianglin and Wang, Jiacheng and Hu, Rose Qingyang and Kim, Dong In and Al-Dhahir, Naofal and Wymeersch, Henk and Zhao, Nan},
  journal={IEEE Trans. Netw. Sci. Eng.},
  title={Intelligent Flexible Position Antenna Systems for Networking: A Survey},
  year={2026},
  volume={13},
  number={},
  pages={3105-3126}}

@ARTICLE{4,
  author={Ding, Zhiguo and Schober, Robert and Vincent Poor, H.},
  journal={IEEE Trans. Commun.},
  title={Flexible-Antenna Systems: A Pinching-Antenna Perspective},
  year={2025},
  volume={73},
  number={10},
  month = oct,
  pages={9236-9253},
  doi={10.1109/TCOMM.2025.3555866}}

@ARTICLE{5,
  author={Ouyang, Chongjun and Wang, Zhaolin and Liu, Yuanwei and Ding, Zhiguo},
  journal={IEEE Trans. Commun.},
  title={Rate Region of {ISAC} for Pinching-Antenna Systems},
  year={2026},
  volume={74},
  number={},
  pages={5849-5866},
  doi={10.1109/TCOMM.2026.3668164}}

@ARTICLE{6,
  author={Liu, Yuanwei and Wang, Zhaolin and Mu, Xidong and Ouyang, Chongjun and Xu, Xiaoxia and Ding, Zhiguo},
  journal={IEEE Commun. Mag.},
  title={Pinching-Antenna Systems: Architecture Designs, Opportunities, and Outlook},
  year={2026},
  volume={64},
  number={1},
  month = jan,
  pages={190-196},
  doi={10.1109/MCOM.001.2500037}}

@ARTICLE{7,
  author={Li, Maolin and Shu, Feng and Yang, Tingting and Zheng, Qinghe and Zhou, Fuhui and Wu, Yongpeng},
  journal={IEEE Trans. Mob. Comput.},
  title={{DQN}-Enabled Joint Pinching Antenna Array Partitioning and Beamforming for Secure {ISAC} Systems},
  note = {early access, Apr. 9, 2026, doi:  10.1109/TMC.2026.3682695}
  }

@ARTICLE{8,
  author={Li, Haochen and Zhong, Ruikang and Pan, Zhiwen and Dong, Chao and Lei, Jiayi and Liu, Yuanwei},
  journal={IEEE Trans. Wireless Commun.},
  title={Pinching Antenna Systems for Integrated Sensing and Communications},
  year={2026},
  volume={25},
  number={},
  pages={13416-13429},
  doi={10.1109/TWC.2026.3668822}}

@ARTICLE{9,
  author={Liu, Peng and Fei, Zesong and Hua, Meng and Chen, Guangji and Wang, Xinyi and Liu, Ruiqi},
  journal={IEEE Trans. Wireless Commun.},
  title={Wireless Powered {MEC} Systems via Discrete Pinching Antennas: {TDMA} Versus {NOMA}},
  year={2026},
  volume={25},
  number={},
  pages={12034-12049},
  doi={10.1109/TWC.2026.3662288}}

@ARTICLE{10,
  author={Liu, Yongxia and Xiao, Jian and Xie, Wenwu and Zhou, Wei and Xu, Hongbo and Cao, Kunrui and Yang, Liang},
  journal={IEEE Internet of Things J.},
  title={Pinching-Antenna-Assisted Distributed Integrated Sensing and Communication Systems},
  year={2026},
  volume={13},
  number={6},
  month = mar,
  pages={11723-11734},
  doi={10.1109/JIOT.2025.3645770}}

@ARTICLE{11,
  author={Jiang, Hao and Wang, Zhaolin and Liu, Yuanwei and Nallanathan, Arumugam and Ding, Zhiguo},
  journal={IEEE J. Sel. Areas Commun.},
  title={Pinching Antenna System {(PASS)} Enhanced Covert Communications: Against Warden via Sensing},
  year={2026},
  volume={44},
  number={},
  pages={4254-4270},
  doi={10.1109/JSAC.2026.3671247}}

@ARTICLE{12,
  author={Hu, Yanglin and Zhang, Tiankui and Xu, Xiaoxia and Liu, Yuanwei},
  journal={IEEE Trans. Veh. Technol.},
  title={Pinching Antenna-Enabled Integrated Sensing and Communication for Low-Altitude {UAV}},
  year={2026},
  volume={75},
  number={6},
  month = jun,
  pages={11728-11733},
  doi={10.1109/TVT.2025.3641978}}

@misc{13,
      title={Pinching-Antenna Enabled Multicell Wireless Systems},
      author={Yunshu Chen and Qing Xue and Meng Hua and Bingpeng Zhou and Shaodan Ma},
      year={2026},
      eprint={2606.12888},
      note = {arXiv:2606.12888},
}

@ARTICLE{14,
  author={Ouyang, Chongjun and Jiang, Hao and Wang, Zhaolin and Liu, Yuanwei and Ding, Zhiguo},
  journal={IEEE Trans. Commun.},
  title={Uplink and Downlink Communications in Segmented Waveguide-Enabled Pinching-Antenna Systems {(SWANs)}},
  year={2026},
  volume={74},
  number={},
  pages={3688-3703},
  doi={10.1109/TCOMM.2026.3653885}}

\end{document}